\documentclass[%
 reprint,
 amsmath,amssymb,
 aps,
]{revtex4-2}

\usepackage{float}
\usepackage{graphicx}
\usepackage{dcolumn}
\usepackage{bm}

\begin{document}

\preprint{APS/123-QED}

\title{Rapid Transition to Remote Instruction of Physics Labs During Spring 2020: Instructor Perspectives}%

\author{Alexandra Werth$^{1,2}$}
\email[]{alexandra.werth@colorado.edu}
\author{Jessica R. Hoehn$^{1,2}$}
\author{Kristin Oliver$^{1,2}$}
\author{Michael F. J. Fox$^{1,2,3}$}
\author{H. J. Lewandowski$^{1,2}$}
\affiliation{$^{1}$Department of Physics, University of Colorado, Boulder, Colorado 80309, USA}
\affiliation{$^{2}$JILA, National Institute of Standards and Technology and University of Colorado, Boulder, Colorado 80309, USA}
\affiliation{$^3$ Department of Physics, Imperial College London, Prince Consort Road, London SW7 2AZ, UK}

\date{\today}

\begin{abstract}
Laboratory courses are an important part of the undergraduate physics curriculum. During physics labs, students can engage in authentic, hands-on experimental practices, which can prepare them for graduate school, research laboratories, and jobs in industry. Due to the COVID-19 pandemic in spring 2020, colleges and universities across the world rapidly transitioned to teaching labs remotely. In this work, we report results from a survey of physics lab instructors on how they adapted their courses in the transition to emergency remote teaching. We found that the instructors who responded to the survey faced numerous challenges when transitioning their classes to remote instruction, particularly in providing students with a similar experience to the in-person labs. In addition, we identified common themes in the instructors' responses including changing learning goals of the courses to be more concept-focused, reducing group work due to equity and technological concerns, and using a variety simulation tools, as well as report on factors that the instructors hoped to continue once they have returned to in-person instruction.

\end{abstract}

\maketitle


\section{\label{sec:Intro}Introduction}

As the COVID-19 pandemic began in the spring of 2020, instructors at colleges and universities worked quickly to move classes and activities to be conducted remotely. The context was unprecedented. Many students were suddenly forced to leave their campus homes, and they faced loss of employment, as well as health concerns for themselves and their family, all while navigating a new modality of learning. Likewise, instructors had limited time to determine new activities, struggled with ethical considerations of remote instruction, and had to learn how to use new technologies, all while handling the impact of COVID-19 on their own personal lives~\cite{RAPIDreport, Brewe}. 

To understand what happened during this switch to emergency remote teaching, including successes and challenges, we developed an online survey to gather information from lab instructors after the spring 2020 semester. This ``instructor survey" was completed by over 100 physics laboratory instructors, mostly in the United States. The survey contained both closed- and open-response questions, which asked instructors about their experience transitioning to remote lab instruction. We previously reported on the initial findings of this survey in a report posted to the arXiv preprint server in July 2020 in order to disseminate the relevant information as quickly as possible to the community~(see Ref. \cite{RAPIDreport}). In this initial report, we found that the transition presented particular challenges for laboratory course instructors, whom often rely on hands-on activities in a complex, collaborative environment involving various technical equipment to help their students learn experimental physics. 

In this work, we expand on the initial quantitative findings using a in-depth analysis of the survey responses and further discuss themes and impacts these findings have on remote instruction of labs beyond the COVID-19 pandemic. Here, when we use the term remote labs we include all continued instruction of a course that was considered a lab course prior to the rapid transition to remote work, and in which the instructor and all students were no longer present at the same location~\footnote{Past distinctions between remote and virtual labs have been made in past work \cite{Ma}. However, due to the forced, rapid transition, we include all courses as to not exclude any courses that do not meet the exact definitions. It is important to note that in spring 2020 most instructors had a few days to weeks to transition their classes to remote instruction, and most preventative measures such as masks and easy access to testing were not available.}.

This study uses a mixed-methods approach---combining statistical analysis on closed-response data with qualitative analysis on rich, open-response data---to identify common themes, challenges, and successes that instructors experienced. These findings can inform both future laboratory instruction and continued research on labs. Although the exact nature of the rapid transition was unprecedented and unlikely to occur regularly, due to the increased availability of technology to teach remotely, many schools are exploring remote alternatives when there are disruptions to teaching due to natural disasters such as fires, hurricanes, and snow--all of which would result in the need to rapidly adapt labs online for short periods of time. Likewise, as demands for online education have grown, so too are our needs to study these environments and determine effective practices in remote instruction. Although many of the choices made by lab instructors during spring 2020 derived from necessity and overwhelming constraints, we were surprised by the number of survey respondents who discussed the successes and things they hoped to continue practicing beyond the pandemic circumstances. In addition, challenges faced by instructors, particularly key lab elements that were missed, such as group work and hands-on experiments, may spur a renewed emphasis on these aspects during in-person labs. Together, these ideas motivated us to answer the following research questions:
\begin{enumerate}
    \item [RQ1.] What motivated the instructors when choosing how to implement a remote lab?
    \item[RQ2.] What challenges did instructors face while implementing and teaching their remote lab courses?
    \item[RQ3.] What strategies did instructors find successful for remote labs? 
    \item[RQ4.] How can the transition to remote instruction inform lab course design for both in-person and remote labs in the future?
\end{enumerate}

The results from this survey can be used to help motivate education researchers to further study the opportunities and limitations of different lab environments and remote strategies. We also hope the presenting of different approaches to remote labs will increase instructors' knowledge of creative practices that could be used for lab courses both during an emergency and outside of such an event to increase opportunities for students generally, as well as those with limited access to in-person instruction.

We begin this work by presenting relevant background on research studying virtual and remote labs prior to the COVID-19 pandemic in addition to contemporaneous studies on the impact COVID-19 has had on physics and lab education (Section~\ref{sec:Background}). In Section~\ref{sec:Methodology}, we provide the methodology for our analysis including our survey administration and design, analysis methods, and qualifications of our results. We structure the results and discussion section of this work, in Section~\ref{sec:Results}, around a number of components that were important: motivating factors, challenges faced, and successes for lab instructors. We follow this by discussing overarching themes that we found from the survey responses. In Section~\ref{sec:beyond}, we draw conclusions reflecting on numerous approaches used by instructors during the transition to remote labs and the impact they may have on future education practices.

\section{\label{sec:Background} Background}

Previous research on the effectiveness of remote lab work has resulted in inconclusive findings, with strong advocates for both traditional hands-on labs and non-traditional approaches~\cite{Ma,foreman,Corter,brinson}. The differences in opinion on remote labs are often attributed to differences in learning goals and objectives between instructors and assessment tools. For example, proponents of hands-on labs often value design skills and social interaction~\cite{Ma}, while proponents of remote labs often value learning content and theory~\cite{brinson}. Other possible benefits of remote lab experiences include providing more flexibility~\cite{SitholeEtal2020} and increasing accessibility for students who are part-time, have disabilities, or have caring responsibilities~\cite{ColwellScanlonCooper2002}.
 
Unlike these past studies that considered intentionally designed remote labs taught by instructors with prior experience navigating an online teaching environment, our work focuses on the unforeseen, urgent, and stressful transition to remote learning due to the pandemic. One of the first decisions instructors had to make was whether to teach synchronously or asynchronously. Synchronous online classes would allow courses to more closely resemble the in-person experience, but could create inequitable classroom experiences for students struggling with technological limitations, new personal responsibilities, or other issues during the pandemic, such as being in a different time zone. Since the start of the pandemic, there have been several studies~\cite{WilcoxVignal2020, Guo2020, FoxEtal2021, PhysRevPhysEducRes.17.010117, RosenKelly2011, LeblondHicks2020, Sahu, educsci10100291, jessicaspaper, PhysRevPhysEducRes.17.020130} that looked at the impact of these types of decisions on physics and STEM classes, and even among these studies there are contradictory findings that speak to the complexity, and the highly context specific nature, of these decisions---a common theme we saw throughout our work.  

For example, in the case of synchronous versus asynchronous instruction, a study by Wilcox and Vignal suggests that there was no difference in student perceived effectiveness for synchronous versus asynchronous lecture formats in their survey population~\cite{WilcoxVignal2020}. However, in a study by Guo~\cite{Guo2020}, which looked at a single physics SCALE-UP~\cite{SCALEUP} style class, they found students who attended the synchronous sessions had an average test grade drop from pre-pandemic of 3.5 percentage points, while students who did not attend had a drop of 14.5 percentage points. In addition, the survey showed that students who did not attend the synchronous sessions found the course more difficult and felt they spent more time on the class than those who attended~\cite{Guo2020}.

In a comparative study of the impact of remote physics lab instruction on student views about experimental physics including over 3200 students, Fox et. al. found that there was no difference in student overall scores on the Colorado Learning Attitudes about Science Survey for Experimental Physics (E-CLASS)~\cite{ECLASS} when comparing courses from both spring and fall 2020 with the same courses in spring and fall 2019~\cite{FoxEtal2021}. Likewise, a study by Rosen and Kelly prior to the pandemic, found that there were no differences in students' epistemological beliefs about experimental physics between the in-person and the online lab~\cite{RosenKelly2011}. However, there were significant differences related to views of socialization; students taking in-person physics laboratories tended to value socialization more than students taking the course online. In another comparative study, Klein et. al. investigated how physics students perceived the sudden shift to online learning during the pandemic. They administered a questionnaire to 578 physics students from five universities in Germany, Austria, and Croatia, and they found that students who collected their own data using real equipment, as opposed to being given data or collecting data using simulations, felt that they gained more experimental skills~\cite{PhysRevPhysEducRes.17.010117}. 

These previous studies show that there is still much disagreement in the field when it comes to best practices and benefits of remote lab courses--whether during a pandemic or not. While remote labs can provide increased flexibility and access for students, they \emph{may} have negative learning outcomes in terms of skill development and socialization. In contrast to these studies, our goal is to highlight multiple approaches to remote lab instruction as described by the respondents to the instructor survey and describe challenges and successes from the instructors' perspective.

\section{\label{sec:Methodology} Methodology}
\subsection{Survey design}
 
\begin{figure}[h]
    \centering
    \includegraphics[scale=0.4]{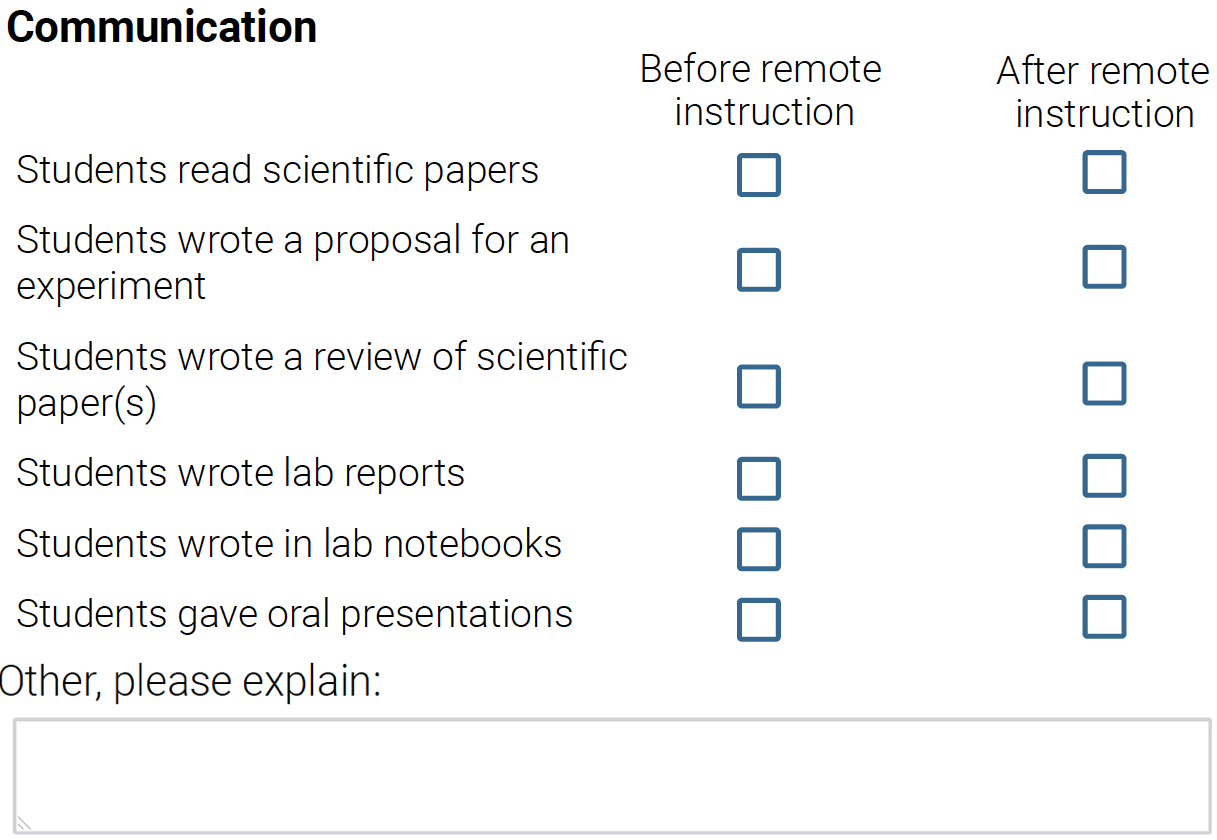}
    \caption{An example question from the instructor survey, where instructors could check boxes to indicate the activities that happened in their courses before and/or after the transition to remote learning.}
    \label{fig:examplequestion}
\end{figure}

The survey was divided into two main sections (Tab. \ref{tab:table1}). First, we asked closed response items where instructors indicated changes that occurred in the course from before to after the transition to remote instruction (Tab. \ref{tab:table1}, Question 1). These closed responses consisted of topics relating to lab structure and activities, course learning goals, student choices, equipment and technology resources, and scientific communication. Additional questions within these categories were added to capture the activities that may be unique to remote labs (e.g., using video conferencing tools). After each of the sections, instructors were given an open ``other" option to describe any additional items that were not captured by the closed-response options, the inclusion of which was motivated by the fact that we had limited knowledge of what instructors were doing given the fast, emergency nature of the transition. An example of a set of questions probing student communication is shown in Figure \ref{fig:examplequestion}. The second half of the survey was comprised of a mix of closed- and open-response questions asking about motivations, challenges, and successes of the remote class (Tab. \ref{tab:table1}, Questions 2-9). For example, instructors were asked to ``Describe the successful aspects of your remote lab class." All of the questions on the survey were optional.

\begin{table*}[]
\caption{\label{tab:table1}%
List and description of questions in the instructor survey.}
\begin{ruledtabular}
\begin{tabular}{ll}
\textbf{Questions} &
  \textbf{Question description} \\ \hline
{\begin{tabular}[c]{@{}l@{}}Question 1: How would you describe the \\ activities in your lab course before and after \\ transitioning to remote instruction? Please \\ select all options that apply.\end{tabular}} &
  \begin{tabular}[c]{@{}l@{}}A list of 42 activities, goals, instructional techniques where \\ instructors have the option to check whether it was done before \\ and/or after the pandemic with the option to write-in “other” \\ activities that were not captured in the closed-responses. These \\ include questions on lab activities and structure, learning goals, \\ student choices, communication, and equipment and  \\ technological resources. \end{tabular} \\ \hline
{\begin{tabular}[c]{@{}l@{}}Question 2: Have you done anything special for \\ your laboratory course this semester that you \\ would like to tell us about?\end{tabular}} &
  Open-ended question \\ \hline
{\begin{tabular}[c]{@{}l@{}}Question 3: When deciding how to teach during \\ the remote instruction portion of the lab, I \\ chose the approach that...\end{tabular}} &
  \begin{tabular}[c]{@{}l@{}}Includes six question statements with 5-point Likert scale options \\ to rate agreement on a scale of strongly disagree to strongly agree.\\ Also includes a option to write-in other factors.\end{tabular} \\ \hline
{\begin{tabular}[c]{@{}l@{}}Question 4: When teaching the remote lab, a \\ challenge I encountered was…\end{tabular}} &
  \begin{tabular}[c]{@{}l@{}}Includes eight question statements with 5-point Likert scale \\ options to rate agreement on a scale of strongly disagree to \\ strongly agree. Also includes open-response question to elaborate \\ on any major challenges faced.\end{tabular} \\ \hline
{\begin{tabular}[c]{@{}l@{}}Question 5: If you were to teach this course\\ remotely again, what would you do differently? \\ Select all that apply.\end{tabular}} &
  \begin{tabular}[c]{@{}l@{}}Check-all-that-apply style question with eight options such as \\ changing lab activities or using difference technology resources. \\ Includes an write-in “other" option.\end{tabular} \\ \hline
{\begin{tabular}[c]{@{}l@{}}Question 6: Describe the successful aspects of \\ your remote lab class.\end{tabular}} &
  Open-ended question \\ \hline
{\begin{tabular}[c]{@{}l@{}}Question 7: What resource(s) or support did \\ you have that was helpful in transitioning to\\ remote instruction?\end{tabular}} &
  Open-ended question \\ \hline
{\begin{tabular}[c]{@{}l@{}}Question 8: What resources or support would \\ have made the remote instruction experience \\ better for you?\end{tabular}} &
  Open-ended question \\ \hline
{\begin{tabular}[c]{@{}l@{}}Question 9: Is there anything else you would \\ like to share about your remote lab class?\end{tabular}} &
  Open-ended question
\end{tabular}
\end{ruledtabular}
\end{table*}

\subsection{Data collection}

Survey volunteers were recruited through professional listservs related to laboratory instruction, as well as through an email to instructors currently administering the E-CLASS~\cite{ECLASS} in their courses. The emails included a link to the survey, which was administered via Qualtrics beginning on April 30th, 2020, with the majority of responses from the instructors being received within the following 2 weeks. Due to the recruitment method, instructors who use E-CLASS represent ~20\% of the survey population. E-CLASS users are particularly interested in formative assessment of their course along the dimension of student epistemologies and attitudes around experimental physics and, therefore, may not be a representative sample of physics lab instructors as a whole.

The survey was completed for 129 courses by 106 unique instructors. A majority of the courses represented in the survey came from 4-year colleges (55\%). Approximately 8\% of the responses were about courses at 2-year colleges, 5\% at Master's granting institutions, and 32\% at PhD granting institutions. Most of the responses came from institutions in the United States (93\%) with 60\% of those being private not-for-profit institutions and 19\% being minority serving institutions. From all the the responses, 61\% of courses were first-year (introductory) labs and 39\% were beyond-first-year labs. Approximately 30\% of the labs were for primarily non-physics/engineering majors, 60\% were for primarily physics and engineering majors, and 10\% to a mixture of majors. Most respondents switched to remote teaching part way through the term, though 17\% of the courses were remote for the entire term (typically from quarter/trimester systems). It is important to note that none of the questions on the survey forced a response; yet, about half of the instructors who completed the survey gave lengthy open-responses that ranged from a few sentences to multiple paragraphs detailing their experiences. We mention this because it is unusual for surveys like these to be filled out so completely; we posit that instructors were eager to share what they did and went through during the semester.

\subsection{Analysis methods}
    \subsubsection{Quantitative methods}
The first set of questions asked instructors to ``Describe the activities in your lab course before and after transitioning to remote instruction," where they were then given a list of activities that might have been part of their course and two possible check boxes representing ``Before remote instruction" and ``After remote instruction" (an example question is shown in Figure \ref{fig:examplequestion}). We calculated the total number of courses that had a given activity for instruction before the transition and the total number of courses that used that activity after the transition to remote labs. We report the uncertainty in these totals using the binomial confidence interval with a 95\% confidence level, where $n$ was 129, the total number of courses. The ``before" and ``after" responses were compared to identify significant changes based on the calculated uncertainty (overlap in the 95\% binomial confidence interval). We report the total number of courses that used the various activities after the transition to remote instruction; in addition, we indicate the direction of the changes in activities between before and after (i.e., more, less, or the same) and the statistical significance to highlight the global trends of our survey population. Thus, we do not not make claims about shifts in activities for individual courses. For questions about learning goals and group work, we present Sankey diagrams, which are flow diagrams in which the width of the lines is proportional to the number of courses represented, to visualize how the nature of individual courses changed from before to after the transition. In addition, we compare differences in the responses between first-year and beyond-first-year courses and 2-year, 4-year, and Ph.D./Masters granting institutions using a Mann-Whitney U test with a $p<0.5$.   

Finally, we asked instructors to rank their agreement to statements about their motivations for the approach they chose, as well as the challenges they encountered.  We used a 5-point Likert scale (from ``strongly agree'' to ``strongly disagree'') for these questions. We treated these data as interval data and assigned a number to each response as follows:  ``Strongly disagree" = 0, ``Disagree" = 1, and ``Neutral" = 2, ``Agree" = 3, and ``Strongly agree" = 4. From this scheme, we calculated means for the responses, with the uncertainty given as the standard error.

  \subsubsection{Qualitative methods}\label{sec:Methodology:qual}
There were several open-response questions on the survey (as shown in Table \ref{tab:table1}). To analyze responses to these questions, we developed two codebooks (summarized in Appendix \ref{app:A}). First, we started with an \emph{a priori} codebook based on the categories of questions asked on the survey as a whole. Many of these main codes also have subcodes, which were created from the closed-response choices of the survey.  Additional subcodes were added during the coding process as emergent codes. These emergent codes were created through a collaborative coding process. AW and KO independently coded a subset of the instructor open-response data (11 courses in total). The percent agreement between the two raters on these 11 responses was 97\%. We report percent agreement instead of Cohen’s Kappa because the large number of subcodes, 99, along with the low prevalence of individual codes across the small data set, can result in unreliable Kappa values~\cite{Gwet2002}. After establishing inter-rater reliability, the entirety of the data set was coded using the first codebook. All additional emergent codes added after the initial inter-rater reliability were discussed and agreed upon by the research team.

As the successes identified by the instructors were critical to answering two of our research questions (RQ3 and RQ4, see Sec.~\ref{sec:Intro}), we wanted to understand these in more detail. Therefore, we developed a second codebook using the open-ended responses that had been coded as \emph{Success} in the first codebook. This second codebook was developed using only emergent coding and captured what instructors found to be successful, their metrics of success, qualifiers (e.g., \emph{at least some} of the students enjoyed...), and things the instructors said they would continue using when they transitioned back to in-person instruction. The ``what was successful" and ``metrics of success" both had subcodes (16 and 23 respectively) that captured nuances of what the instructors considered successful and why. For example, an instructor wrote \emph{``Since the goal was primarily to explore physics concepts, I think the use of simulations helped us to still meet that goal."} In this case, the ``what was successful" were simulations and the ``metric of success" was students learning physics concepts. JH and KO separately coded 20 responses that were coded as ``Success" using the first codebook. These 20 responses were not used in the second codebook creation process. The percent agreement between the two raters for the 20 responses was found to be 93\%. Both codebooks are available in Appendix \ref{app:A}.

\subsubsection{Limitations}
In developing the survey, we were aware that the closed-response options we provided would unlikely be able to capture the full breadth of experiences faced by the instructors. However, through the analysis of the open-response questions, as described in Section \ref{sec:Methodology:qual}, we were able to supplement the closed-response data with instructor provided responses (which are included in the relevant figures in Appendix~\ref{app:B}). The prevalence of these responses should be considered in the context that some were prompted and others unprompted, and so they may be considered a demonstration of existence.

The wording and interpretation of the survey questions was not validated beyond the research team due to the time-sensitive nature of the research. Therefore, we cannot be certain that all instructors interpreted the questions in the way that we intended. However, the responses to the open-response questions reported in this work were consistent with our intentions when writing the survey. Furthermore, we focus primarily on these open-responses in our analysis and thus instructors' interpretation of the questions, while still a limitation, do not impede our ability to draw meaningful conclusions from the results.

Another limitation of this study is that some instructors may not have had the time, energy, or ability to fill out an online survey due to increased stress and responsibilities due to the pandemic. Access to technology, having a quiet space to work, attending to family responsibilities, and dealing with both mental and physical healthcare were challenges not only for students, but for instructors as well. This is of particular concern given that the pandemic has had a disproportionate impact on women and those from marginalized groups~\cite{minello2020pandemic,doi:10.1177/00027642211003140, doi:10.1080/00918369.2020.1868184, 10.1093/geronb/gbaa117}. We did not collect the demographic information of the instructors surveyed, but we suspect that the sample of instructors might be biased in this way because women and marginalized groups carried a disproportionate burden of stress and responsibilities during this pandemic time. When drawing our conclusions in this study, we remain sensitive to these missing perspectives and hope that the results are interpreted with this in consideration. 

In addition, we did not ask instructors to report on the race, ethnicity, gender, or socioeconomic status of their student populations because most instructors do not have easy access to this information. The context and constraints faced by instructors vary based, in part, on the student population (e.g., instructors who teach students who are majority low income may have had a different set of considerations to take into account when determining how to structure a remote lab class); however, we did not want to further burden instructors with finding this information or needing to guess the demographic make-up of their students. This represents a limitation of closed-response portion of our study where we cannot compare the constraints faced by our instructors to other outside factors.

\section{\label{sec:Results} Results and Discussion}
We first present the motivations and challenges concerning the transition to remote lab instruction in order to answer RQ1, ``\emph{What motivated the instructors when choosing how to implement a remote lab?}" and RQ2, ``\emph{What challenges did instructors face while implementing and teaching their remote lab courses?}". Throughout our discussion, we present results regarding the lab structure, technological choices, and activities to document the tools instructors used to address these challenges. Next, we present results of strategies instructors felt were successful in this switch to remote labs. Here, we describe not only what instructors identified as successes, but the metrics they used to determine what was a success answering RQ3, ``\emph{What strategies did instructors find successful for remote labs?}''. There were many different aspects of the remote labs that were successful and many metrics of success used across the courses in our dataset. The implementation of remote labs was idiosyncratic and highly context specific. That being said, there are a few common themes that emerged during the analysis. We use these common themes to answer our final research question, RQ4 ``\emph{How can the transition to remote instruction inform lab course design for both in-person and remote labs in the future?}".

\subsection{\label{sec:Motivations and challenges} Motivations and challenges}

\subsubsection{Motivations}

\begin{figure*}[ht]
    \centering
    \includegraphics[scale=0.6]{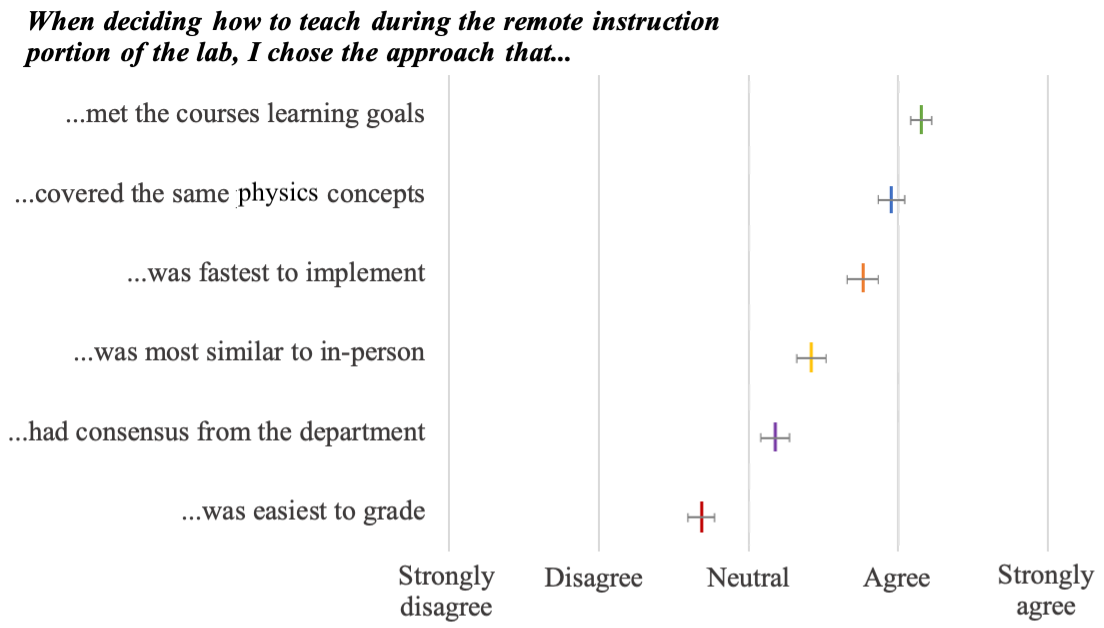}
    \caption{Instructors were asked to ``Rank how much you agree with the following statements.'' We show the mean response from 121 survey responses and the error, which represents one standard error of the mean. We calculated the mean by assigning a response of ``Strongly disagree" = 0, ``Disagree" = 1, and ``Neutral" = 2, ``Agree" = 3, and ``Strongly agree" = 4.}
    \label{fig:motivation}
\end{figure*}

We found that, although the instructors described a range of motivations, most were driven by the desire to meet the course learning goals and to cover the same concepts as before remote instruction (see Figure \ref{fig:motivation}). Some instructors listed following departmental consensus and ease of grading as reasons for making their decisions, but overall these were not the primary motivators when designing the remote version of the course. Another motivation that was not represented in the closed response options, but discussed in the the open responses was ensuring the remote course was equitable and accessible (i.e., all students in the class had access to the resources they needed to learn). For example, one instructor explained they \emph{``had to find things that worked that students could do without buying stuff.''} For another, their main motivation was to ensure the well-being of their students: 

\begin{quote}
    \emph{I prioritized mental health by holding mental health check ins at the beginning of every class period.  This really helped the class to create a community and also reinforced with the students that I valued them as people first.  I have found that students will work harder and learn more if you care for them as a whole person.}
\end{quote}

These motivations were likely dependent on a variety of factors including departmental culture, student population demographics, and even local spread of COVID-19, which varied dramatically between geographic locations. For example, in April 2020 New York City had one of the highest rates of local spread in all of the United States with ~10,000 new cases on April 10th alone, whereas the entire state of Florida had 10 times fewer new cases on the same day. Although these factors were not captured in the survey, what we did see was a large variety of primary motivations. 

\begin{figure*}[ht]
    \centering
    \includegraphics[scale=0.95]{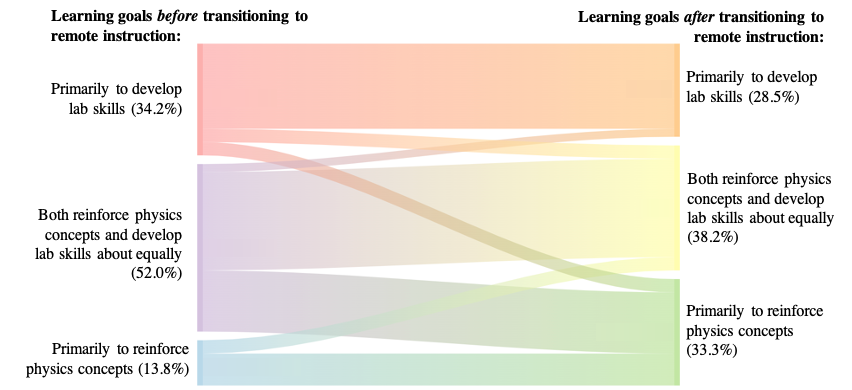}
    \caption{Sankey diagram showing the change in learning goals of the instructors who completed the instructor survey from before (left side of plot) to after (right side of plot) remote instruction. The lines represent the direction of change from before to after and the width of the line is proportional to the number of courses that reported that type of transition. }
    \label{fig:LearningGoals}
\end{figure*}

The majority of instructors indicated that they chose the instructional strategies to try to meet the learning goals for the course. When asked about the broad learning goals of the class (developing skills, reinforcing concepts, or a mixture of both) for both before and after the transition, instructors reported switching more towards emphasizing physics concepts rather than lab skills. This was particularly true for instructors that, prior to the emergency remote instruction, had the course focus on both concepts and skills about equally (Figure \ref{fig:LearningGoals}). 

From the closed-response data, we cannot comment on the specific ``skills" and ``concepts" that various courses focused on. That is, it is difficult to know if the shift of courses that emphasized both concepts and skills prior to remote instruction to primarily learning physics concepts was because the ``skills" learning goals were centered around using hands-on equipment (e.g., soldering), which students were unable to do remotely. However, based on the open-responses, this was the case for some of the instructors. For example, one instructor from a small, beyond-first-year course whose goal before and after the transition to remote instruction was to reinforce both skills and concepts equally wrote, \emph{``Teaching lab skills involving hand-on use of equipment was not possible"} and indicated that after the transition, they primarily reinforced concepts. Another instructor from a Ph.D. granting institution teaching a small, beyond-first-year course whose goal before and after the transition to remote instruction was to primarily reinforce skills wrote, 

\begin{quote}
\emph{One of the three course goals involves developing students' ability to use the tools and techniques that experimentalists [sic] use in the lab. This is pretty much impossible remotely.}
\end{quote}

However, as we see in Figure \ref{fig:LearningGoals}, the majority of courses with primary learning goals associated with skills maintained those learning goals after the transition, with many people finding creative ways to focus on laboratory skills in the remote classes. Another survey respondent, an instructor from a Master's granting institution teaching a beyond-first-year, small laboratory course for physics and engineering majors said, 

\begin{quote}
\emph{Even though no lab work occurred after remote instruction began, students had to rely on their notebooks and previous data collection to complete required oral presentations and written reports, both considered part of `lab skills' (i.e., experimental physics skills).}
\end{quote}

It is important to note that the above two quotes were both from beyond-first-year courses. Traditionally, beyond-first-year and first-year courses have very different learning goals with beyond-first-year courses more heavily emphasizing lab skill development~\cite{AAPT, PhysRevPhysEducRes.16.020162}. We see in Table \ref{tab:learninggoals_BYF_FY} that this was the case prior to the transition to online instruction--26.6\% of first-year courses compared to 11.8\% of beyond-first-year courses focused on primarily physics concepts. However, there was a 6.5 times increase in the number of beyond-first-year courses primarily emphasizing physics concepts after the transition to remote instruction compared to only a 1.9 times increase in first-year-courses having this learning goal. This implies that the beyond-first-year courses struggled more with maintaining skill based learning goals than first-year courses. This may be due to the fact that beyond-first-year courses often rely on more complex laboratory equipment and experimental set-ups that could not be easily translated into the remote learning environment.

\begin{table}[h!]
\caption{\label{tab:learninggoals_BYF_FY}%
Transition in learning goals before and after the transition to remote instruction for first-year and beyond-first-year courses}
\begin{ruledtabular}
\begin{tabular}{lcc}
Learning Goal                                                              & \% before & \% after  \\ \hline
~                                                           
\bf{First-year courses} (n = 73) & &  \\
Primarily physics concepts                                                              & 21.9 & 41.1                                     \\
Primarily lab skills                                                          & 28.8 & 24.7                                     \\
Both about equally                                                         & 49.3 & 32.8                                     \\
\hline
\bf{Beyond-first-year courses} (n = 50) & &  \\
Primarily physics concepts                                                              & 4.0 & 26.0                                    \\
Primarily lab skills                                                          & 44.0 & 36.0                                     \\
Both about equally                                                         & 56.0 & 42.0                                     \\
\hline
\end{tabular}
\end{ruledtabular}
\end{table}

Related to this result, the only survey questions about motivations that had statistically significantly different responses (using a Mann-Whitney U test with $p < 0.05$) between first-year and beyond-first-year courses were ``...met the courses learning goals" and ``covered the same physics concepts." In both cases, first-year courses were more likely to strongly agree that these were motivating factors compared to beyond-first-year courses. A further discussion on the implications of the transition in learning goals can be found in Section \ref{sec:Learning goals_Dis}.

\subsubsection{Challenges}

\begin{figure*}[ht]
    \centering
    \includegraphics[scale=0.6]{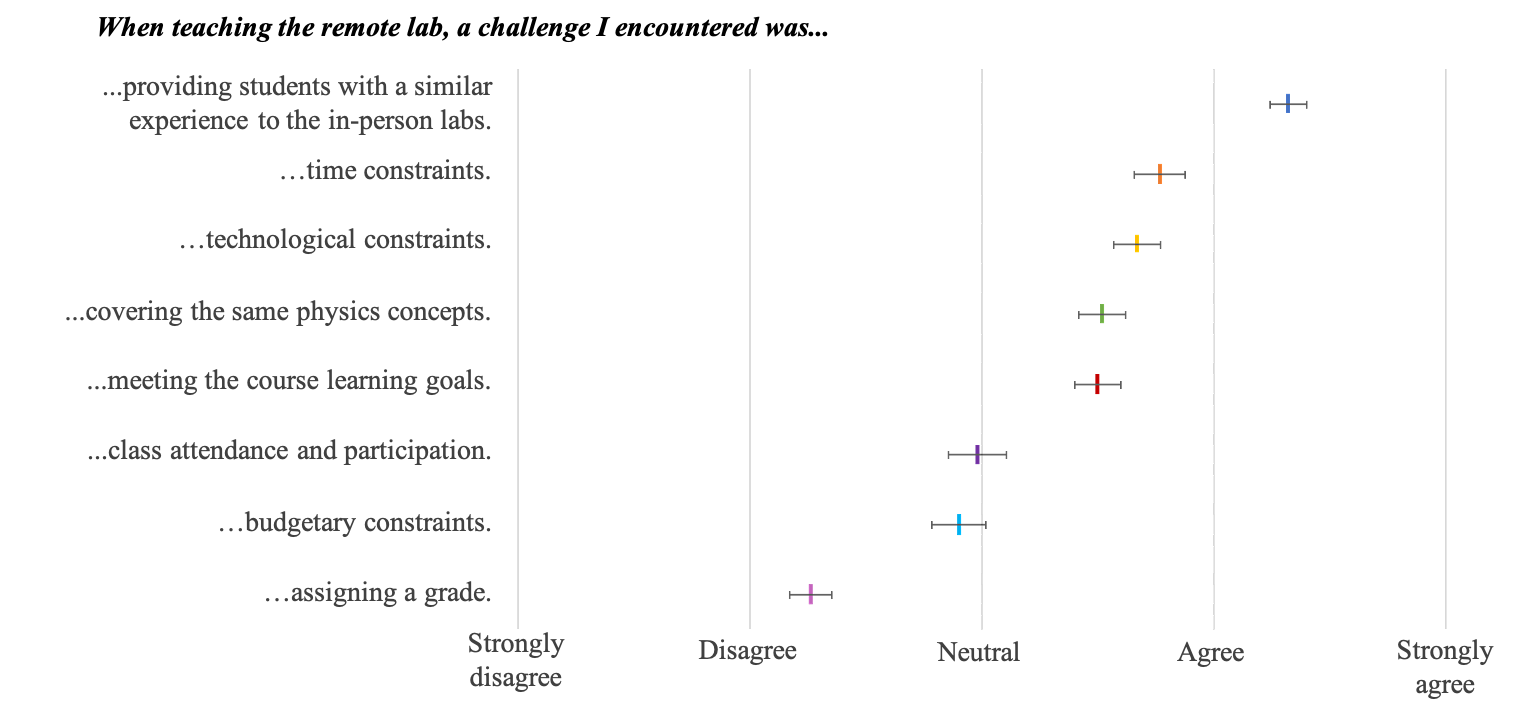}
    \caption{Instructors were asked to ``Rank how much you agree with the following statements.'' We show the mean response from 111 survey responses and the error which represents one standard error of the mean. We calculated the mean by assigning a response of ``Strongly disagree" = 0, ``Disagree" = 1, and ``Neutral" = 2, ``Agree" = 3, and ``Strongly agree" = 4.}
    \label{fig:challenges}
\end{figure*}

The most common reported challenge instructors faced was making the remote class as similar to the in-person version as possible. Instructors also cited time and technology constraints as major challenges (Figure \ref{fig:challenges}). Grading was not indicated as a major challenge, perhaps because a large number of institutions switched to pass/fail grading schemes or because many instructors were encouraged to be more lenient with their grading in the emergency remote situation. Responses to the statements on class attendance/participation and budget were somewhat polarized (which is not represented by the mean shown in Figure \ref{fig:challenges}). Two challenges that were commonly cited in the open-responses, but not captured in the closed-responses were student engagement (22 of 129 courses) and missing hands-on components of the lab (18 courses). As an example, an instructor wrote \emph{``motivation suffered when we switched from hands-on work to all computer-based work."} Other challenges captured in the open-responses included lack of resources (15 courses), difficulty fostering student collaboration and group work (15 courses), personal factors for the instructor and students, such as family responsibilities (7 courses), too high of a workload for students or instructors (5 courses), and maintaining equity for the students (3 courses).

Many instructors navigated not being able to use hands-on lab equipment by turning to new technologies. However, it was clear from the survey that with new technology comes new challenges. Thirty-eight instructors discussed a myriad of challenges they faced when using new technology. A common theme in those responses, and the second most common challenge faced by the instructors (Figure \ref{fig:challenges}), was that time was a major limitation when choosing and using various equipment and online technologies. One instructor, who taught a large course for non-physics/engineering majors, did not have time during the spring 2020 semester to develop experiments using cell phone applications as a measurement system, but was hoping to develop this curriculum over the summer of 2020: 

\begin{quote}
    \emph{We are transitioning this summer to hands-on labs using cell phones and basic materials.  I'm excited that this will work out well.  Time to research how to teach online is the most important tool.  Time to collaborate with other universities.  Time to collect equipment for students.  There was no way to do this in the spring, teach other courses online (again with no support) and develop labs.}
\end{quote}

Another instructor for a large, introductory level course for non-physics/engineering majors wrote that it took \emph{``considerable extra time"} to develop new curriculum that \emph{``aligned with the course goals"} particularly because they had to use a variety of different technological solutions depending on the specific lab or task they wanted students to do. They used \emph{``simulations, provided data, home-made videos and photos, YouTube videos"} and had multiple iterations with feedback from other staff and teaching assistants to coordinate the new labs for the large class. The same instructor noted that due to time constraints, navigating the new technologies was not only a challenge for them but also for the students: 

\begin{quote}
    \emph{Gauging how long labs would take was difficult. Students reported spending 4, 5, 6, 7 hours on a nominally 3-hour lab. Students reported being unable to find 3 continuous hours at any point during the week to work on it. We adjusted as the course went on and feel like the time was less of an issue at the end of the quarter.}
\end{quote}

There were no statistically significant differences (using a Mann-Whitney U test) in the responses to the survey questions on challenges between first-year and beyond-first-year courses. However, it is possible that their may be difference based on other factors such as type of institution (Ph.D./masters granting, 4-year college, or 2-year college). It is difficult to draw conclusions on statistical differences of responses based on these factors due to the low numbers of responses from instructors of 2-year colleges. We do note that all of the 2-year college courses reported attendance/participation as a challenge, whereas many 4-year and PhD/Masters granting institutions did not.

\subsection{\label{sec:Successes} Successes and metrics of success}

Beyond categorizing the strategies used by instructors for remote labs and understanding the vast array of challenges faced by the instructors, we also wanted to know what they deemed successful and how that might impact future lab courses, and thus answer our final two research questions. The data we present below come from a wide range of different instructional environments. While each context has its own unique challenges, and there is clearly not a single solution, we hope that we can illustrate a range of what instructors thought worked well and what indicators they relied upon to measure that success.

\subsubsection{\label{sec:Metrics of success} Metrics of success} 
In determining what was successful, we must first understand what metrics of success the instructors themselves considered given their individual contexts, values, teaching approaches, and goals. For example, the success of a given strategy or course may be measured by student affect (i.e., did students enjoy the course?), addressing learning goals of the course (whether preserved from the in-person course or novel to remote teaching), or simply completing the term during the pandemic.

\begin{figure*}[ht]
    \centering
    \includegraphics[scale=0.75]{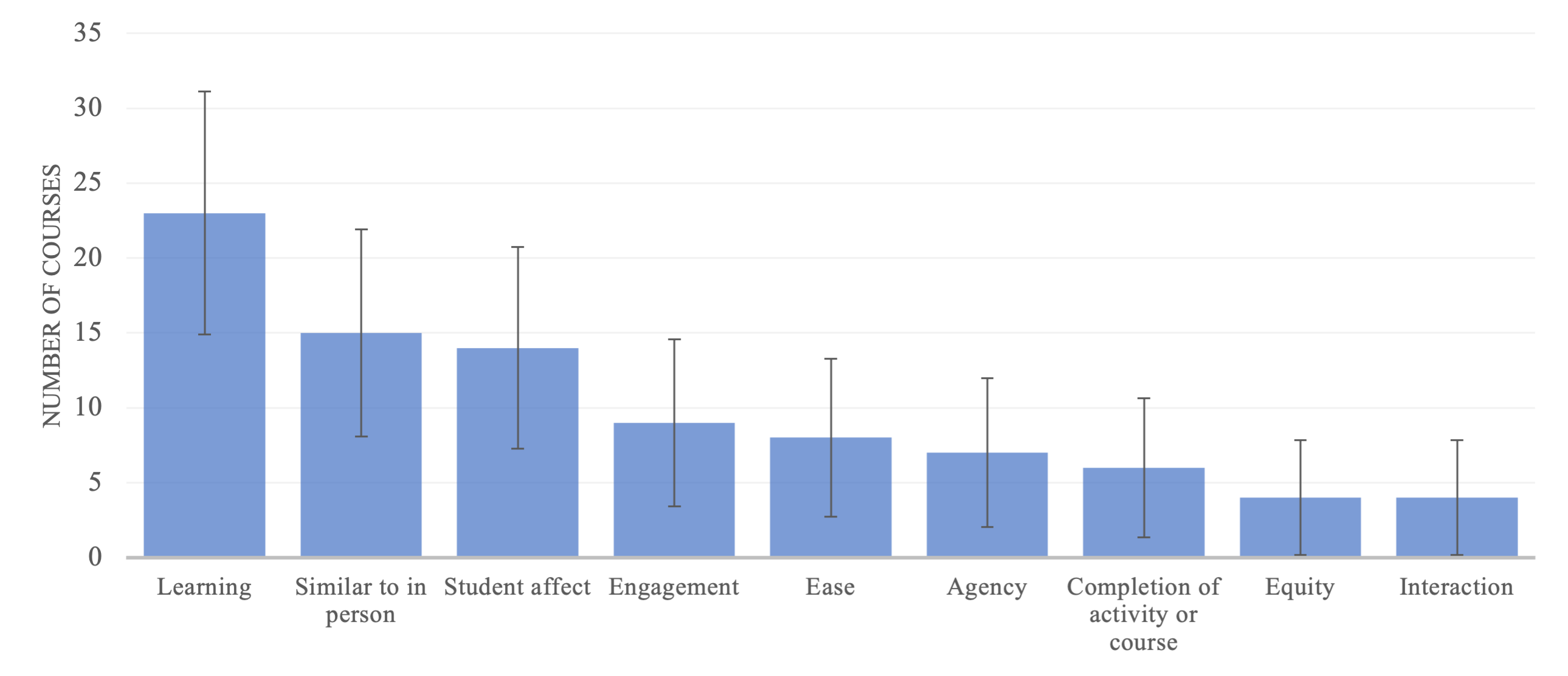}
    \caption{The metrics of success referenced by instructors in the open responses of the survey. The error bars were calculated using the 95\% binomial confidence interval with $n=102$, the number of references to any metric of success by the instructors.}
    \label{fig:Metricsofsuccess}
\end{figure*}

We did not ask instructors directly about their metrics of success on the survey; however, during the coding analysis, there were 102 unprompted references to various metrics of success by the instructors. Figure \ref{fig:Metricsofsuccess} shows that learning, which was defined meeting the course learning goals, was the most commonly reported metric of success---reported for 23 of the courses represented in the survey. This was closely followed by the metric of achieving a similar experience to the in-person lab and by the metric of student affect. Other metrics of success mentioned by three or fewer instructors, and not represented in Figure \ref{fig:Metricsofsuccess}, were having continuity during the transition to remote instruction, student creativity, quality of student work, building community, and effective science communication through writing and presentations.

\subsubsection{Successes}

\begin{figure*}[ht]
    \centering
    \includegraphics[scale=0.75]{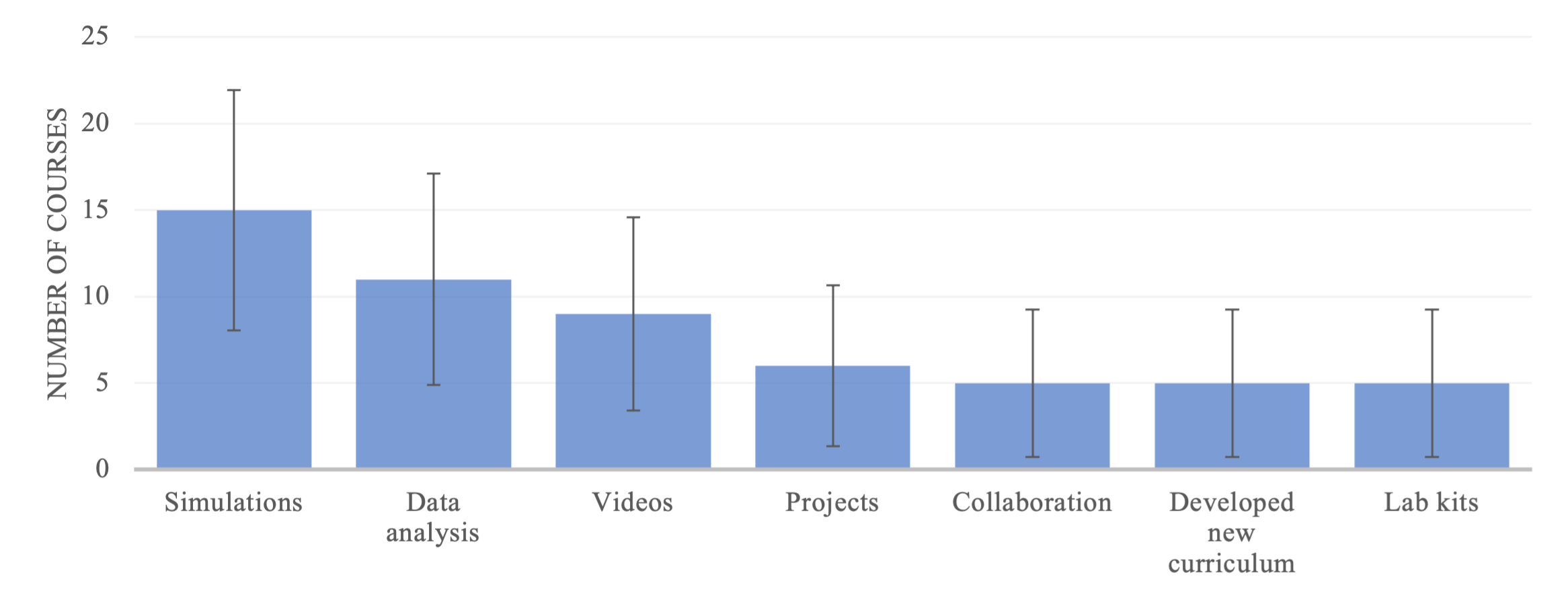}
    \caption{The successes referenced by instructors in the open responses of the survey. The error bars were calculated using the 95\% binomial confidence interval with $n=91$, number of references to any success by the instructors. As noted in Appendix \ref{app:A}, Tab. \ref{tab:table3}, there were 22 success subcodes--we present only the subcodes that are statistically above zero (using the 95\% binomial confidence interval) in this figure.}
    \label{fig:Successes}
\end{figure*}

We can also look at the aspects of the course that instructors found to be successful for achieving their goals. In an open response question of the survey, we asked instructors to ``Describe the successful aspects of your remote lab class." In addition to the responses to this question, we coded all successes referenced within any of the other open responses. There were over 25 unique items that instructors found to be successful in the transition to remote lab instruction. The most commonly referenced success, shown in Figure \ref{fig:Successes}, was in using simulations. Instructors of 15 different courses used simulations successfully across many metrics of success, including remaining similar to the in-person experience (5 courses) and achieving learning goals of the course (4 courses). In addition, simulations were described as successful in a variety of different course types, see Table \ref{tab:simulationssuccessful}. An instructor from a medium-sized, introductory lab for non-physics/engineering majors wrote, \emph{``The use of the photoelectric effect and blackbody PhET simulations was very successful."} Likewise, an instructor from a small, beyond-first-year electronics course for physics majors wrote, 

\begin{quote}
    \emph{In an electronics lab, switching to SPICE simulations for remote instruction actually worked pretty well. It came late enough in the semester that students had already developed decent electronics lab skills (test and measurement skills, breadboarding, grounding, debugging, etc.). Though not ideal, students got an extra skill (SPICE) and were able to still pick up the main physics of new circuit components.}
\end{quote} 

Beyond the successful aspects indicated in Figure \ref{fig:Successes}, other instructors found electronic assignments (e.g., online lab notebooks or textbooks), using household items or cell phones, learning management systems, small group check-ins, synchronous virtual meetings, student presentations, journal club, and student direction of instructor data collection to all be successful for their courses.

Because there was often a disconnect between when an instructor talked about metrics of success and the successful item itself, we are not generally able to show which activities were the most successful given specific metrics of success. There are, however, several specific cases where instructors found certain course activities to be successful because of the metrics of success that were important to them. We present a few of these specific examples here.

One instructor of a small, introductory-level course measured their students' success via their level of engagement. The instructor described students participating in an experiment \emph{``to determine the functional form for the force produced on a suspended point charge by a charged rod,"} and explained that the students collaborated via video conference to discuss their experimental design. The group of students then provided the instructor with a set of instructions and treated the instructor as a \emph{``lab assistant."} The instructor said, 

\begin{quote}
    \emph{This was a real experience to them in every way I was able to detect. They were engaged (even more deeply because of the considerable increase in messaging between the groups and myself due to the remote class) with every phase of the experiment.}
\end{quote}

For this instructor, the necessary changes in methods of collaboration due to the pandemic actually led to an increase in student engagement~\cite{jessicaspaper}.

As another example, an instructor of a small, beyond-first-year course found that shifting the lab goals towards data analysis and computation allowed them to be successful because it allowed for student learning that would not necessarily have been a part of their in person course. The instructor said,

\begin{quote}
   \emph{We moved to mostly more sophisticated computer modeling of real data, some of which was student data, some was instructor data. So students did have to develop their analysis procedures and troubleshoot their code.}
\end{quote}

Their goal for the remote transition had been to \emph{``do what could be done really well in the remote learning environment."} Ultimately, the instructor felt that their course had been successful because of an interaction they had with a student after the semester was over. The instructor said, 

\begin{quote}
    \emph{One student talked to me after the semester was over and said he was finally starting to get over some of his anxiety about programming. For this student, the push toward computation, data analysis, and modeling was a very good opportunity to grow.}
\end{quote} 

For this instructor, the shift in their goals allowed students to gain a valuable learning experience that would not necessarily have taken place without remote learning. 

Finally, an instructor of a medium sized, introductory-level course found the creation of online lab videos to be successful, measured by the level of student interactions. This instructor brought lab equipment home and made videos of the lab experiments, which were then posted to YouTube. Students worked with these homemade videos in small groups. The instructor found that these videos worked well and allowed students to take data. In response to the open-ended question about success on the survey, this instructor said \emph{``In lieu of 3-4 students standing around a table, we used Zoom breakout rooms. Discussion was very active within each room." }

Eight instructors found some of the elements to be so successful that they plan on keeping them once moving back to in-person instruction. For example, one instructor created three new lab experiments that could be done remotely and commented that, in the future, they would be \emph{``suitable for students with certain disabilities."} Another instructor changed the lab curriculum to be more open-ended and project based~\cite{jessicaspaper}, 

\begin{quote}
\emph{Previous quarters the labs were more cookbook and focused on making sure students saw a particular result. The remote version was more open ended and more like a guided research project. We relied on weekly group meetings with students to help them figure out how to proceed through their project. The level of student engagement was much higher in the remote format. Students were much more engaged in problem solving and making meaningful decisions about what to do and how to do it.  We intend to retain much of the remote format once we go back to in-lab operations. }   
\end{quote} 

Given that many of the instructors expressed struggles with student engagement in the remote lab, it is interesting to see that a more open-ended project led to \emph{increased} engagement despite the remote environment.

\begin{table}[b]
\caption{\label{tab:simulationssuccessful}%
Types of courses that found simulations to be successful. }
\begin{ruledtabular}
\begin{tabular}{cc}
Course Type                                                               & Number of Courses \\ \hline
Introductory                                                              & 13                                    \\
Beyond-first-year                                                         & 2                                     \\ \hline
Small (1-25 students)                                                                    & 7                                     \\
Medium (25-100 students)                                                                    & 5                                     \\
Large ($>$100 students)                                                                     & 3                                     \\ \hline
Physics/engineering majors      & 6                                     \\
Non-physics/engineering majors & 9                                    
\end{tabular}
\end{ruledtabular}
\end{table}

It is important to acknowledge that there were far more challenges than successes noted by the instructors responding to the survey. However, we were inspired by the optimism from many of the instructors' responses and hope that it also informs future physics lab teaching practices both in-person and remote. In addition, we can look at common themes in our analyses to further understand the implications for future lab instruction.

\subsection{Major components of labs that were distinct in the remote setting}

\subsubsection{\label{sec:Simulations_Dis} Success of simulation use}
Some lab courses switched to using simulations as sources of data collection, making measurements, and learning physics content. Past research, conducted prior to the pandemic, found that simulations were useful for reinforcing physics concepts~\cite{Jimoyiannis2001, Perkins2006, Adams2010}. This is particularly true as some simulations have been developed to address specific and common student difficulties~\cite{Guangtian2011}. We saw in the survey responses that a larger percentage of courses emphasizing physics concepts as their primary learning goal after the transition to remote instruction used PhET simulations~\cite{phet} (47.5\%) than courses whose focus was learning lab skills during remote instruction (20.5\%).

However, multiple instructors in the survey discussed the usefulness of other simulations for developing lab skills. Eleven instructors mentioned using simulations beyond PhET including: Fritzing~\cite{fritzing}, KET~\cite{KET}, MultisimLive~\cite{multisim}, oPhysics~\cite{ophysics}, SPICE~\cite{Nagel:M382}, MATLAB’s Simulink~\cite{matlab}, The Physics Aviary~\cite{physicsaviary}, and students coding their own simulations. Some simulations allow students to interact with models of physical phenomena via their computers or smartphones and engage in authentic decision making, data collection, and troubleshooting practices. These simulations tend to have larger parameter spaces for students to explore such as Pivot Interactives\cite{pivot}--a hybrid of simulation and video analysis, where real experiments have been filmed with a variety of different parameter selections. These allow students to explore the real-world parameter space and, using overlaid measurement tools, perform measurements from the videos. An instructor from a small, beyond-first-year course for engineering and physics majors called simulations \emph{``valuable"} and wrote, \emph{``I might use them as part of a class even with in-person learning."}

In addition, many electronics labs found circuit simulations such as SPICE, MATLAB's Simulink~\cite{matlab}, or MultisimLive~\cite{multisim} particularly useful because students were able to build and model `real' circuits (with non-idealized performance). Since these tools are commonly used in industry, this also meant that students could still have an authentic lab experience and develop important lab skills. One instructor wrote, \emph{``Given the original design of the lab activities, a combination of Fritzing~\cite{fritzing} and MultisimLive~\cite{multisim} allowed students to practice many of the skills I had already planned to address.''} While some instructors noted that simple simulations may not be able to replicate the complex aspects of performing experiments in real life, the example above of using Fritzing~\cite{fritzing} may emulate more what working on circuit design is like for professionals, than compared to using simpler simulations.

\subsubsection{\label{sec:Learning goals_Dis} Shift of learning goals}

After the transition to remote instruction, courses that previously emphasized both concepts and skills about equally tended to move towards learning physics concepts (Figure \ref{fig:LearningGoals}). This aligns with the literature, which finds that many proponents of online labs value learning physics concepts (i.e., content and theory) where proponents of hands-on labs often value design skills and collaborative skills~\cite{Ma,foreman,Corter,brinson}. Perhaps the online environment maybe better at meeting the learning goals associated with learning physics concepts in contrast to lab skills. 

Given the extenuating circumstances, pivoting the learning goals of a lab course to focus more on concepts may have been a reasonable, productive, and effective solution. However, a past study has shown that physics laboratory courses that focus specifically on developing lab skills promote more expert-like beliefs about the nature of experimental physics than courses that focus either on reinforcing physics concepts or on both goals~\cite{Wilcox2017}. As we see from the instructor survey, the majority of courses with primary learning goals associated with skills were able to maintain those learning goals after the transition, with many instructors developing creative approaches to still focus on laboratory skills in the remote classes. 

The ability to maintain a focus on experimental skills during remote instruction depends on the resources available to students and instructors, as well as on what skills are considered important. The obvious challenge associated with remote lab instruction is the potential absence of hands-on interaction with measurement devices and experimental apparatus, particularly if the lab requires sophisticated and expensive equipment. Some classes were able to continue hands-on experimentation in spring 2020 by sending equipment home to students or having students use resources from home, including the use of smartphone applications as measurement devices (Appendix \ref{app:A}, Figure \ref{fig:LabActivities}). Additional instructors indicated that they were \emph{``looking into lab kits to be sent to student's homes"} for future remote terms. However, many of these home lab kits do not allow for opportunity to learn how to work with more complex measurement apparatus--an important aspect of many beyond-first-year physics labs~\cite{AAPT}. 

Another common learning goal for labs includes developing skills associated with data and uncertainty analysis~\cite{PollardMay21}. To do this, many instructors sent students data that they had collected previously or that they generated for the purposes of the course. Alternatively, instructors asked students to review data from scientific publications or publicly available data sets, since the development of data analysis skills does not necessarily require students to collect their own data. However, using previously collected or open-source data may diminish student understanding of how the data was acquired and how it should be interpreted. Some instructors overcame this challenge by having students control equipment remotely, watch videos of the instructor(s) take data, or even have students provide the instructor with directions with how to collect the data.  An instructor from a small, introductory lab did just this: 

\begin{quote}
    \emph{Students actively instructing me in the conduct of and [sic] experiment that has video footage to analyze does not appear to be different for many labs goals than for them to do it themselves, at a surprisingly high level.}
\end{quote}

Still, the instructor felt that their students missed, 

\begin{quote}
    \emph{...actual manipulative skills that would come from handling the equipment, and some agency and executive function skills that are not exercised because the setting constrains choices to a smaller range than students would face when confronted with a document and equipment.}
\end{quote}

It is clear that some necessary physics lab skills are challenging to replicate completely in a remote environment. Nonetheless, some skills were less affected by the transition to remote instruction than others. As we see in Appendix \ref{app:A}, Figure \ref{fig:Communication}, courses focusing on the development of scientific writing, reading, and presentation skills---especially writing lab reports, giving oral presentations, reading scientific papers, writing experimental proposals, and writing reviews of scientific papers---were able to continue doing this after the transition.

\subsubsection{\label{sec:Collaboration_Dis} Collaboration in an online environment}

\begin{figure*}[ht]
    \centering
    \includegraphics[scale=0.4]{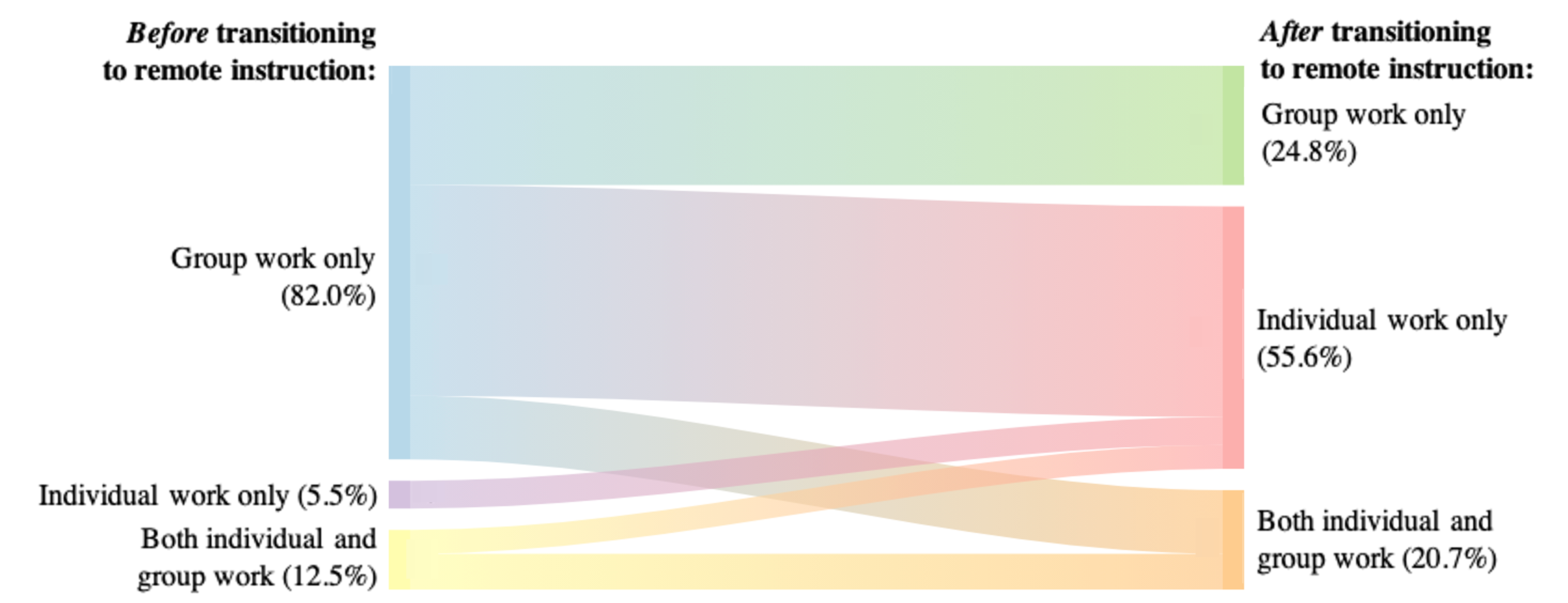}
    \caption{Sankey diagram showing the change in courses designed for group work versus individual work during lab from before (left side of plot) to after (right side of plot) remote instruction. The lines represent the direction of change from before to after and the width of the line is proportional to the number of courses who reported that type of transition.}
    \label{fig:groupwork}
\end{figure*}

Collaboration is often an essential part of labs. The AAPT Recommendations for the Undergraduate Physics Laboratory Curriculum~\cite{AAPT} suggest that one of the goals for students in physics labs should be to develop ``interpersonal communication skills" through ``teamwork and collaboration." In addition, as research and business is increasingly conducted in a global environment, many believe that it is essential for students to be prepared to engage in effective collaborations remotely within diverse groups~\cite{Borrego,Fredrick,Goltz}.  However, in the survey, some instructors expressed that they were \emph{``concerned about the mechanisms of group work with the rapid transition to online"} so they switched to mostly individual work. Due to logistical concerns, other instructors gave students the option to opt-out of group work. One instructor wrote, \emph{``After remote transition, students could work in groups or individually.  I did not structure it either way. Before we transitioned, I structured it as group work."} And the move to individual work was not isolated to only a few instructors; in a closed-response question, we asked the instructors whether students ``worked in groups with other students" or ``worked individually" before and after the transition to remote instruction. Figure \ref{fig:groupwork} shows that before the transition to remote instruction, 82\% of the courses reported that students only worked in groups. After the transition to remote instruction, only 24.8\% of students exclusively worked in groups, while the percentage of courses where students worked only individually on labs increased to 55.6\%.  

From the open-responses, the most common motivation for this change toward individual work was equity concerns. Due to the sudden nature of the transition, many instructors could not or did not want to require students to attend labs synchronously: 

\begin{quote}
    \emph{We could not require that all the students perform the lab synchronously during the designated lab time. (Some students had limitations on internet access). So we allowed students to complete the lab activity asynchronously within a 30 hr time frame. This has led to some students opting to work alone with the data collection part and engage in group activity only at the report writing stage.}
\end{quote}

These decisions resonate with recent reports on the impact of the COVID-19 pandemic on college students that suggest that being a student emerged as a higher risk factor for loneliness during lockdown than usual~\cite{BU202031} in addition to increased worry and grief~\cite{CONRAD2021117}. Higher levels of social capital and sense of community are significantly associated with lower levels of loneliness~\cite{THOMAS2020103754}. Fostering group work in the online classroom can pose new challenges~\cite{Koh}; nonetheless, it can have overwhelming benefits including increased motivation, creativity, and reflection~\cite{Johnson}--essentials during a time of increased isolation for students.

Five of the instructors found methods for successful student collaboration despite the challenging circumstances. One instructor said that the students continued to engage with the material and \emph{``even more deeply because of the considerable increase in messaging between the groups and myself due to the remote class." }

Small breakout rooms in Zoom seemed to help some with enabling collaboration. One instructor expected group work to be a larger issue in the remote setting, but found that it was not as challenging as expected \emph{``as long as I kept the groups to three students."} Another said, 

\begin{quote}
    \emph{Students could still work productively in groups trying to do sense-making activities---the zoom break out rooms (and my ability to pop-in and pop-out of those rooms to address the problems the students were grappling with worked better than I thought it would).}
\end{quote} 

A fourth instructor wrote, \emph{``Group projects came out fine even given the challenges.  Students all continued to participate at the same level, so no issues there."}

Finally, one instructor noted an increase in the students reaching out for help, \emph{``...student requests for assistance increased compared to traditional instruction.  Students used GroupMe (online chat app used for course) to request Zoom meetings to go over topics."} They noted that facilitating the course became a community effort and everyone in the course, including the students and instructors, had to work \emph{``together to ensure information was accessible to all and was updated in a timely fashion."} 

The examples of successful aspects of remote lab courses seen in our dataset lead us to wonder how this experience may positively (or negatively) influence physics lab education beyond the pandemic. For example, online group work clearly posed a challenge for many instructors and one instructor wrote that they would have liked to have ``resources for how to manage group work online." An increase in accessible resources describing some best practices in online group work could help instructors who are (a) interested in moving their lab courses online or (b) need to quickly switch to online labs in response to volatile weather, future pandemics, or other natural disasters.

\section{Beyond spring 2020 and conclusions \label{sec:beyond}}

As shown in the instructor responses, there are many benefits to in-person labs that are difficult or impossible to replicate in an online environment. Collaborating face-to-face with instructors and peers to troubleshoot technical equipment is just one example of an staple of in-person physics lab and essential to student growth and development as experimental physicists~\cite{AAPT}. Perhaps these missed elements of in-person physics lab instruction will spur a renewed interest from instructors and increase their emphasis during in-person labs. However, it is important to realize, that online education is likely to increase in the future due to various pressures. 

 Online education has grown steadily since the early 2000s due to new technologies, global adoption of the Internet, and a demand for a college-educated workforce~\cite{allen_2009, allen_2007, doi:10.1080/1097198X.2018.1542262, ORTAGUS201747}. Some are postulating that for college teaching and learning, there may be no return to normal since the COVID-19 pandemic will disrupt the notion that courses taught online are significantly worse than in-person learning~\cite{NYT}. It is reasonable to expect online learning in higher education will be in our future and understanding best pedagogical practices in the online environment is essential--particularly for labs. 

Although the rapid, unpredictable nature of this transition led to an extreme set of challenges, some of these issues with moving labs online will likely persist even without time constraints on the implementation of the courses. For example, how can we make remote labs a similar experience to in-person labs -- which often rely on hands-on equipment? And without students in the classroom, how can we teach the necessary lab skills, and do so safely? Perhaps more importantly, should we even attempt to move labs to an online environment for the majority of students? Regardless, if we think most labs should return to being primarily in person, could we make labs more accessible to students with disabilities or to those who do not have access to physical labs, such as at remote locations?  

Through this work, we identified some tools used by instructors, such as focusing on scientific writing skills and using authentic simulation tools, to successfully implement lab-like learning in a remote setting. However, similar to past literature~\cite{Ma,foreman,Corter,brinson, WilcoxVignal2020, Guo2020, FoxEtal2021, PhysRevPhysEducRes.17.010117, RosenKelly2011, LeblondHicks2020, Sahu, educsci10100291}, we found that approaches that worked in some institutional and classroom contexts were not successful in others. To illustrate, an instructor from a small, beyond-first-year class teaching a modern physics lab said, 

\begin{quote}
    \emph{I could imagine a class where experiments are done by the students at home, but given the different life circumstances of students, the class would likely not be an equitable experience.}
\end{quote}

On the other hand, an instructor from another school teaching a medium-sized, introductory lab said, \emph{``I had them measure the focal length of their cell phone camera lens based on the recent paper... Worked well!"} However successful, it is important to note that this solution could lead to inequitable experiences for students if they do not all own a cell phone with a camera. 

Aside from online pedagogy, some instructors used this as an opportunity to try a new curriculum that they hope to bring back to the in-person experience. Five instructors reported that they plan to incorporate simulations into the future in-person experience. In addition, one instructor implemented contract grading: 

\begin{quote}
    \emph{So one day it hit me to try a grading contract, which has really minimized how much I have to formally grade. I give students feedback but they get credit for completion, so the grading burden is a lot smaller on me... I think when we go back to in person I am going to try some sort of hybrid so every single check in doesn't have to be graded, but students must show completion of everything to get some type of minimal grade.}
\end{quote}

Another instructor \emph{``took advantage of the free LabArchives"} and said that \emph{``students gave positive feedback on that so I'm considering switching to e-notebooks next year."} Lastly, as we saw in Section \ref{sec:Successes}, one instructor was able to move from \emph{``cookbook"} experiments that were \emph{``focused on making sure students saw a particular result"} to a \emph{``guided research project"} in the remote version. They found that student engagement actually increased in the remote format and the open-ended format allowed students to be \emph{``engaged in problem solving"} and make \emph{``meaningful decisions"} about the experimental processes.

This work presents the wide range of approaches that instructors employed in spring 2020 in order to teach remote physics lab classes, and demonstrates some of the possible ways to successfully conduct remote labs, as well as some of the common challenges. We encourage future studies to continue analyzing the impact of remote labs on student learning, particularly from the student perspective. More work must be done to investigate the ability to achieve common physics lab learning goals, the impact on student development of professional collaborative skills, and student identity as experimental physicists in online environments.

\begin{acknowledgments}
We thank all the instructors who took the time to share their experiences with us. This work is supported by NSF RAPID Grant No. DUE-2027582 and STROBE National Science Foundation Science Technology Center, Grant No. DMR-1548924.
\end{acknowledgments}

\bibliographystyle{unsrt}
\bibliography{references}

\clearpage

\appendix

\section{Codebooks \label{app:A}}

Here, we provide additional details of our two codebooks described in Section \ref{sec:Methodology}. The first table (Table \ref{tab:table2}) provides a description of the main codes from the primary codebook. The second table (Table \ref{tab:table3}) provides descriptions of the success codes. 

\begin{table*}[ht]
\caption{\label{tab:table2}%
Description of major codes in first codebook.}
\begin{ruledtabular}
\begin{tabular}{ll}
\textbf{Codes}               & \textbf{Code descriptions}                                                         \\ \hline
Lab activities               & \begin{tabular}[c]{@{}l@{}}Includes eleven sub-codes such as watched videos or lab with simulations, that describe the \\ activities done in the course\end{tabular}                                                                                                    \\ \hline
Lab structure                & \begin{tabular}[c]{@{}l@{}}Includes nine sub-codes which describe the structure of the course either before or after the \\ transition to remote instruction. Examples include asynchronous activities and project-based.\end{tabular}                                                                                                                  \\\hline
Learning goals               & \begin{tabular}[c]{@{}l@{}}Coded whenever an instructor discussed their learning goals for the course. This code did not \\ have sub-codes.\end{tabular}                                                                                         \\ \hline
Student choice               & \begin{tabular}[c]{@{}l@{}}Includes ten sub-codes detailing student choices in the course such as working at their own \\ pace or designing a procedure.\end{tabular}                                                                          \\ \hline
Communication                & \begin{tabular}[c]{@{}l@{}}Communication is broken into two categories: (1) General or logistical communication and \\ (2) scientific communication.  General communication includes six sub-codes which categorize \\ the type of communication (e.g., whether it was amongst peers or students with instructors). \\ The scientific communication sub-code has eight sub-codes which represent communication \\ based lab activities (e.g., oral presentations or writing in lab notebooks).\end{tabular} \\ \hline
Equipment and technology     & \begin{tabular}[c]{@{}l@{}}Has 33 sub-codes categorizing the variety of equipment and technological resources used by \\ the instructors. Includes details of specific product/company names.\end{tabular}           \\\hline
Motivating factors           & \begin{tabular}[c]{@{}l@{}}Coded when instructors discuss any motivating factors as to their decisions when choosing \\ how to run the remote course. Does not have any sub-codes.\end{tabular}                                           \\ \hline
Challenges                   & \begin{tabular}[c]{@{}l@{}}Includes fifteen sub-codes such as personal life of the instructor or student, equity, and time to \\ capture the types of challenges  faced by the instructional team and students.\end{tabular}                         \\ \hline
Successes                    & \begin{tabular}[c]{@{}l@{}}Coded whenever instructors discuss a success. These codes were later used to create second \\ codebook (see Appendix \ref{app:A}, Table \ref{tab:table3}).\end{tabular}            \\ \hline
Resources that were helpful  & Coded when discussing helpful resources; no sub-codes. \\\hline
Resources that we would like & \begin{tabular}[c]{@{}l@{}}Resources that would have been helpful but were not available; no subcodes.\end{tabular}     \\\hline
Changes to lab               & \begin{tabular}[c]{@{}l@{}}Describes changes that instructors \emph{did} make or would like to make in the following semester \\ to the remote lab. This code includes seven subcodes which detail what the instructors would \\ have like to change (e.g., course content or lab activities).\end{tabular}                                                     \\\hline
Collaboration                & \begin{tabular}[c]{@{}l@{}}An emergent code for when instructors discussed students working in groups or individually.\end{tabular}               \\\hline
Student engagement           & \begin{tabular}[c]{@{}l@{}}An emergent code for when instructors discussed students' levels of engagement in the remote \\ course.\end{tabular}                                                      \\ \hline
Creativity                   & \begin{tabular}[c]{@{}l@{}}An emergent code for when instructors discussed students thinking creatively or when \\ instructions described creativity as a learning goal.\end{tabular}                      
\end{tabular}
\end{ruledtabular}
\end{table*}

\begin{table*}[]
\caption{\label{tab:table3}%
Description of major codes in second codebook used for ``successes."}
\begin{ruledtabular}
\begin{tabular}{ll}
\textbf{Codes}          & \textbf{Code descriptions}                                                                                                                                                                                                                                                                                                                                                                                                   \\ \hline
What was successful     & \begin{tabular}[c]{@{}l@{}}Includes 22 subcodes such as collaboration or small group check-ins that detail what exactly\\ instructors found successful during the remote lab\end{tabular}                                                                                                                                                                                                                                    \\ \hline
Metrics of success      & \begin{tabular}[c]{@{}l@{}}Includes 15 subcodes which describe why an instructor believed a certain element was successful.\\ For example, one instructor found ``synchronous sessions" to be successful because they had \\ ``good attendance and participation." In this case, synchronous labs would be coded as ``What\\ was successful" and student engagement would be coded as the ``Metric of success."\end{tabular} \\ \hline
Qualifiers              & \begin{tabular}[c]{@{}l@{}}This was recorded to note when an instructor qualified a success. For example, ``I hosted a review\\ on Blackboard Collaborate that was successful, though not as good as in-person." This instructor\\ found the learning management system to be successful, but qualified that it was still not as good\\ as in-person labs.\end{tabular}                                                      \\ \hline
Would use for in-person & \begin{tabular}[c]{@{}l@{}}Coded when an instructor indicated that they would like to carry an activity or practice back into\\ the in-person labs. For example, ``The simulation labs were valuable.  I might use them as part of \\ a class even with in-person learning."\end{tabular}                                                                                                                    
\end{tabular}
\end{ruledtabular} 
\end{table*}

\clearpage
\section{Additional close-response data \label{app:B}}

Here, we provide the data from the closed-response questions on the survey about \emph{what} was done in the labs before and after the transition to remote instruction along with some descriptions highlighting important aspects of the data. In these figures, the y-axis is the number of courses who reported using these activities after switching to remote instruction, except for the ``other" category, which is the number of courses that did ``other things" based on the open responses. The inset plot shows the breakdown of the ``other" category. The green bars represent a statistically significant (when comparing the 95\% binomial confidence intervals) increase in that activity after the transition to remote instruction. The red bars represent a statistically significant (when comparing the 95\% binomial confidence intervals) decrease in that activity after the transition to remote instruction. The solid blue bars represent a change in that activity that was not statistically different from before the transition. We do not know how many instructors used activities in the ``other'' category before the transition, so they are denoted by striped blue bars. All significance and errors bars were calculated using the 95\% binomial confidence interval.

\subsection{\label{sec:Lab activities} Lab structure and activities}
\begin{figure*}[ht]
    \centering
    \includegraphics[scale=0.65]{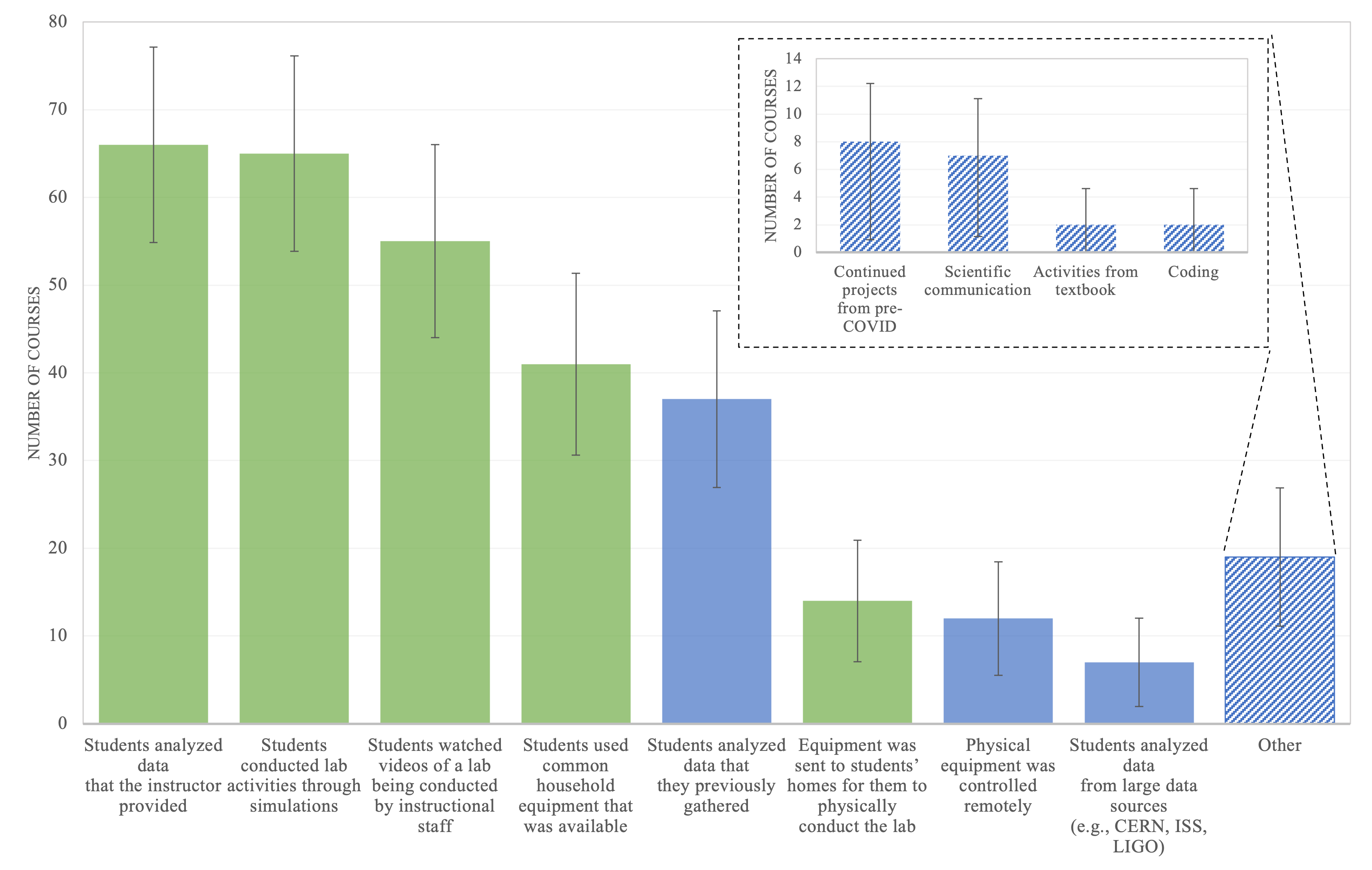}
    \caption{Instructor reported lab activities used after the transition to remote instruction in spring 2020.}
    \label{fig:LabActivities}
\end{figure*}

After the transition to remote instruction, many instructors changed both the lab structure and the activities in their courses. In particular, instructors did not have as many traditional guided labs (74 before the transition to 43 after the transition with a 95\% binomial confidence interval of $\pm$11) and they increased the number of asynchronous lab activities. These asynchronous activities included having students analyze data provided by the instructor and having students use simulations as replacements for the in-person lab activities (Figure \ref{fig:LabActivities}).

There were significant increases (green bars in Fig.\ref{fig:LabActivities}) in activities analyzing instructor provided data, labs conducted through simulations, students watching videos of labs being conducted by instructional staff, students using household equipment to complete lab activities, and equipment being sent from the school to students' homes. In addition, instructors discussed in their open-responses to the survey that they had students continue to work on projects from before the transition and focus on scientific communication (``other" category in Figure~\ref{fig:LabActivities}). Given that many institutions transitioned to remote instruction in the middle of the spring 2020 semester, some classes were able to pivot and extend projects that the students had already started. Others who could not conduct previously planned experiments due to the remote environment, opted to have students write review papers on a scientific topics.

\clearpage

\subsection{\label{sec:Agency} Student choices and self-regulated learning experiences}

Activities in an online learning environment are often very different from conventional lab classes. Many remote classes had fewer or no opportunities for students to interact face-to-face with their instructors and classmates. During asynchronous components of online classes, students are more responsible for their own learning, as they decide when, where, and for how long to work on course activities and assignments; therefore, self-regulated learning behaviors are especially important when taking online courses~\cite{Wang2013, BARNARD20091}. For remote physics labs, this could potentially result in added student decision making about their own analysis methods, lab procedures, troubleshooting experimental apparatus at home, building their own apparatus, and developing their own research questions.   

\begin{figure*}[ht]
    \centering
    \includegraphics[scale=0.7]{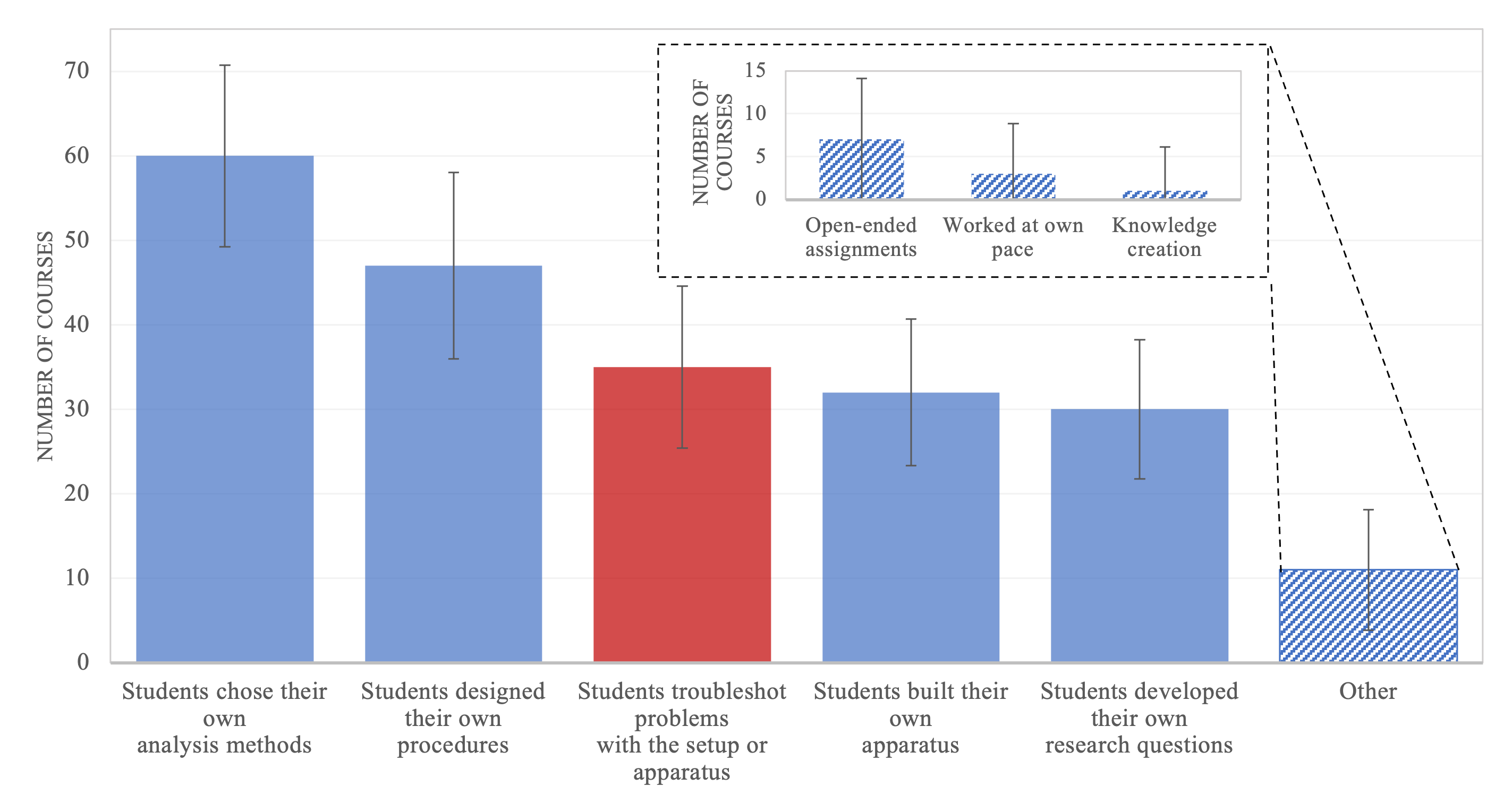}
    \caption{Instructor reported student choice and self-regulated learning opportunities used after the transition to remote instruction in spring 2020.}
    \label{fig:Agency}
\end{figure*}

Figure \ref{fig:Agency} shows that after the transition to remote instruction, many instructors said that their students were able to continue to choose their own analysis method with no significant decrease compared to in-person learning. 

Although the number of students choosing their own analysis methods remained the same from before to after the transition, we saw a significant drop in students troubleshooting problems with experimental setups--likely due to fewer hands-on activities. However, one instructor found that computer modeling of data allowed students to work on their troubleshooting skills without needing hands-on labs. Another instructor considered that in many fields of science, researchers do their science remotely (e.g., astronomers, high-energy physicists) when not physically constructing detectors. Although this instructor does not explicitly mention troubleshooting, it is true that many physicists who work with remote equipment engage in the troubleshooting processes---for example, the Hubble flawed mirror design~\cite{hubble}.

\clearpage

\subsection{\label{sec:Communication} Communication and collaboration}

We asked instructors about engaging with various forms of scientific writing, reading, and presentation before and after the transition to remote instruction.  This scientific communication stayed relatively similar; however, there was a significant drop from 82 to 56 courses (with a 95\% binomial confidence interval of $\pm$11) in which students wrote in lab notebooks. In the open responses, instructors did not report on asking students to engage in any forms of scientific communication outside of what was captured in the closed responses. However, some instructors needed to adapt the technology used to the remote environment such as engaging in oral presentations, literature review, and using Google Docs.

\begin{figure*}[ht]
    \centering
    \includegraphics[scale=0.75]{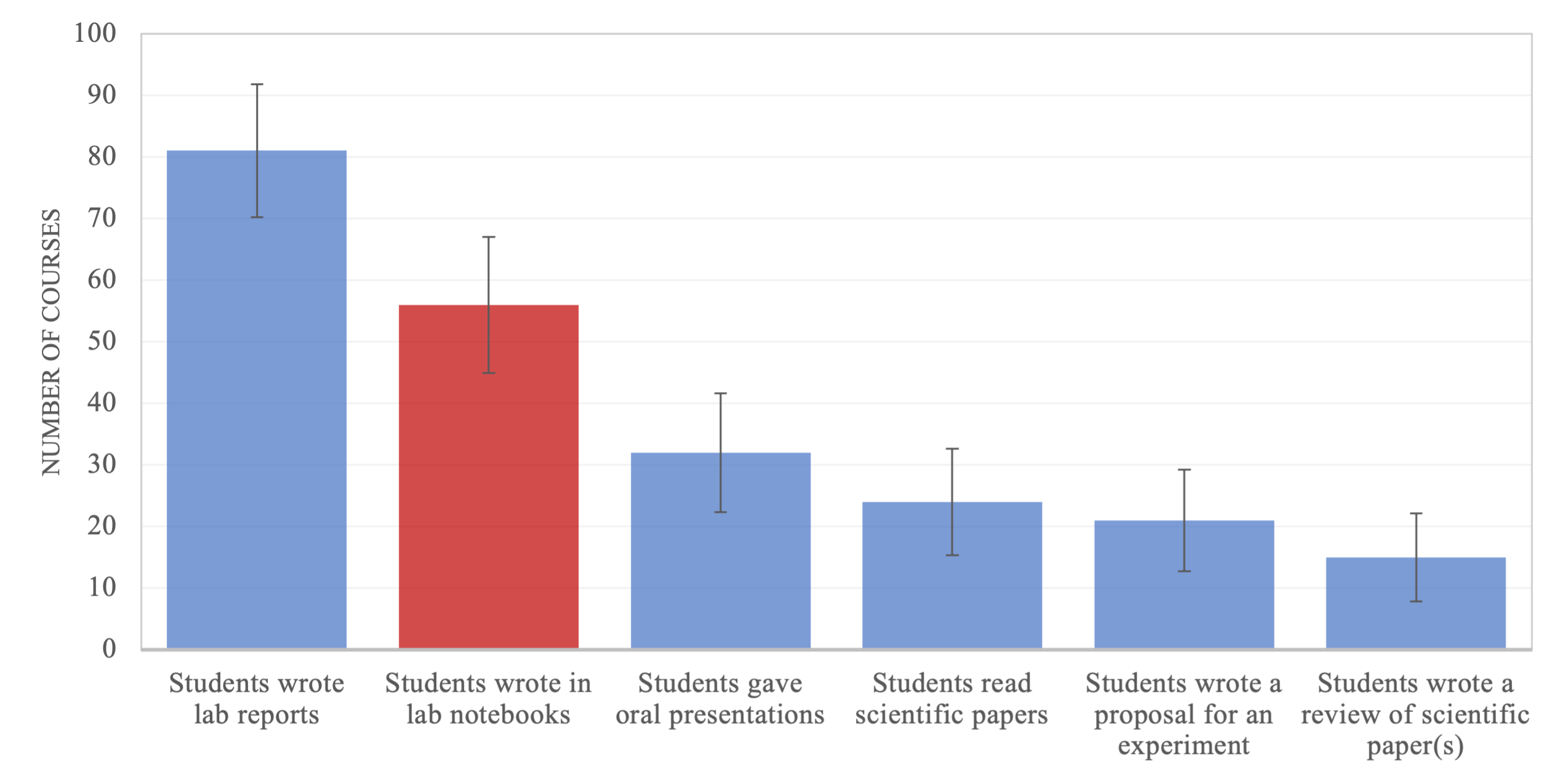}
    \caption{Instructor reported scientific communication activities used after the transition to remote instruction in spring 2020.}
    \label{fig:Communication}
\end{figure*}

\clearpage

\subsection{\label{sec:Technology and equipment} Technology and equipment}

As instruction moved online, most instructors who responded to our survey used video conferencing technology in order to interact with their students and hold their classes. However, beyond regular face-to-face (or, in this case, screen-to-screen) communication, labs often require data analysis and experimental design---aspects that can require creative technological solutions to move online.

\begin{figure*}[ht]
    \centering
    \includegraphics[scale=0.55]{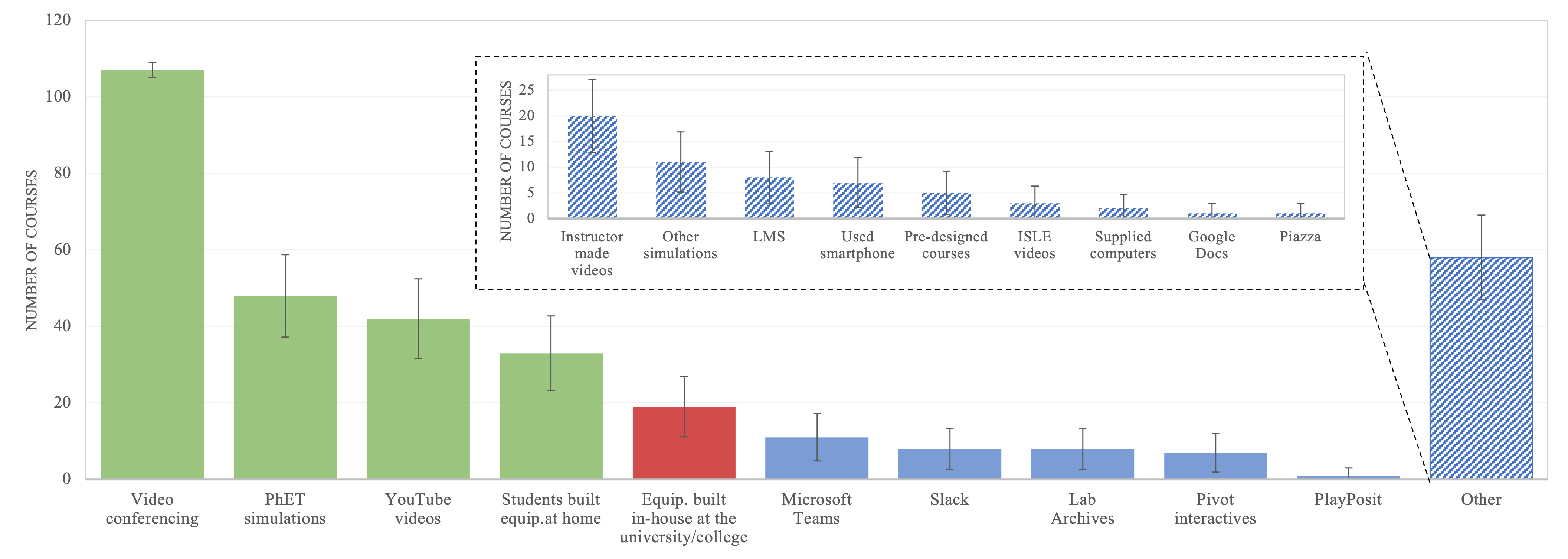}
    \caption{Instructor reported equipment and technology used after the transition to remote instruction in spring 2020. }
    \label{fig:EquipmentandTechnology}
\end{figure*}

After the transition, there was a significant increase in the reliance on PhET simulations, YouTube videos, and students building their own equipment at home (green bars in Figure \ref{fig:EquipmentandTechnology}). There was a significant decrease in students using university-housed equipment (i.e., equipment or experimental set-ups made by the school for students). Additionally, we see in Figure \ref{fig:EquipmentandTechnology} that there were many other diverse technological solutions instructors used that were not captured by the closed-response portion of the survey. In the open responses, 20 instructors mentioned that they created their own videos for students.

\end{document}